\documentclass{aa501}
\usepackage{graphicx}
\usepackage{rotating}
\begin{document}

\newcommand{\as}[2]{$#1''\,\hspace{-1.7mm}.\hspace{.1mm}#2$}
\newcommand{\am}[2]{$#1'\,\hspace{-1.7mm}.\hspace{.0mm}#2$}
\def\approxlt{\lower.2em\hbox{$\buildrel < \over \sim$}}
\def\approxgt{\lower.2em\hbox{$\buildrel > \over \sim$}}
\newcommand{\dgr}{\mbox{$^\circ$}}   
\newcommand{\grd}[2]{\mbox{#1\fdg #2}}
\newcommand{\gsim}{\stackrel{>}{_{\sim}}}
\newcommand{\HI}{\mbox{H\,{\sc i}}}
\newcommand{\HIbf}{\mbox{H\hspace{0.1 em}{\footnotesize \bf I}}}
\newcommand{\HIit}{\mbox{H\hspace{0.1 em}{\footnotesize \it I}}}
\newcommand{\HIsl}{\mbox{H\hspace{0.1 em}{\footnotesize \sl I}}}
\newcommand{\HII}{\mbox{H\,{\sc ii}}}
\newcommand{\IHI}{\mbox{${I}_{HI}$}}
\newcommand{\Jykms}{\mbox{Jy~km~s$^{-1}$}}
\newcommand{\kms}{\mbox{km\,s$^{-1}$}}
\newcommand{\kmsMpc}{\mbox{ km\,s$^{-1}$\,Mpc$^{-1}$}}
\def\lir{{\hbox {$L_{IR}$}}}
\def\lco{{\hbox {$L_{CO}$}}}
\def \ls{\hbox{$L_{\odot}$}}
\newcommand{\LB}{\mbox{$L_{B}$}}
\newcommand{\LBnul}{\mbox{$L_{B}^0$}}
\newcommand{\LBsun}{\mbox{$L_{\odot,B}$}}
\newcommand{\lsim}{\stackrel{<}{_{\sim}}}
\newcommand{\LsunB}{\mbox{$L_{\odot, B}$}}
\newcommand{\LsunMsun}{\mbox{$L_{\odot}$/${M}_{\odot}$}}
\newcommand{\masq}{\mbox{mag~arcsec$^{-2}$}}
\newcommand{\MHI}{\mbox{${M}_{HI}$}}
\newcommand{\MHILB}{\mbox{$M_{HI}/L_B$}}
\newcommand{\MHILBfr}{\mbox{$\frac{{M}_{HI}}{L_{B}}$}}
\def \ms{\hbox{$M_{\odot}$}}
\newcommand{\Msun}{\mbox{${M}_\odot$}}
\newcommand{\MsunLsun}{\mbox{${M}_{\odot}$/$L_{\odot,Bol}$}}
\newcommand{\MsunLBsun}{\mbox{${M}_{\odot}$/$L_{\odot,B}$}}
\newcommand{\MsunLKsun}{\mbox{${M}_{\odot}$/$L_{\odot,K}$}}
\newcommand{\MT}{\mbox{${M}_{ T}$}}
\newcommand{\MTLBnul}{\mbox{${M}_{T}$/$L_{B}^0$}}
\newcommand{\MTLBsun}{\mbox{${M}_{T}$/$L_{\odot,B}$}}
\newcommand{\tis}[2]{$#1^{s}\,\hspace{-1.7mm}.\hspace{.1mm}#2$}
\newcommand{\Vcor}{\mbox{${V}_{0}$}}
\newcommand{\Vhel}{\mbox{$V_{hel}$}}
\newcommand{\VHI}{\mbox{${V}_{HI}$}}
\newcommand{\vrot}{\mbox{$v_{rot}$}}
\def\la{\mathrel{\hbox{\rlap{\hbox{\lower4pt\hbox{$\sim$}}}\hbox{$<$}}}}
\def\ga{\mathrel{\hbox{\rlap{\hbox{\lower4pt\hbox{$\sim$}}}\hbox{$>$}}}}
\newcommand{\eff}{\mbox{Effelsberg}}
\newcommand{\nan}{\mbox{Nan\c{c}ay}}
\newcommand{\pks}{\mbox{Parkes}}
\newcommand{\gb}{\mbox{Green Bank}}

  \title{A neutral hydrogen survey of polar ring galaxies}

  \subtitle{IV. Parkes observations}

  \author{W. van Driel\inst{1},
          F. Combes\inst{2},
          M. Arnaboldi\inst{3},
                \and
          L.S. Sparke\inst{4}
          } 

  \offprints{W. van Driel}

  \institute{GEPI, Observatoire de Paris, Section de Meudon,
             5 place Jules Janssen, F-92195 Meudon, France \\
             \email{wim.vandriel@obspm.fr}
        \and
             LERMA, Observatoire de Paris, 61 avenue de l'Observatoire,
             F-75014 Paris, France \\
             \email{francoise.combes@obspm.fr}
        \and
             Osservatorio Astronomico di Capodimonte, V. Moiariello 16, 
             Napoli 80131,  Italy \\
             \email{magda@cerere.na.astro.it}
        \and
            Astronomy Department, University of Wisconsin-Madison,
            475 N. Charter St., Madison WI 53706, U.S.A. \\
            \email{sparke@uwast.astro.wisc.edu}
             }

    \date{\it Received 12/12/2001 ; accepted 12/2/2002}

  \abstract{{\rm
A total of 33 polar ring galaxies and polar ring galaxy candidates were observed in 
the 21-cm \HI\ line with the 64-m Parkes radio telescope. The objects, selected by 
their optical morphology, are all south of declination --39\dgr\ 
and in only 5 of them \HI\ had been reported previously. 
\HI\ line emission was detected towards 18 objects, though
in 3 cases the detection may be confused by another galaxy in the 
telescope beam, and one is a marginal detection.
Eight objects were detected for the first time in \HI, of which 5 did not have 
previously known redshifts.}
   }

 \authorrunning{W. van Driel et al. }
 \titlerunning{A neutral hydrogen survey of polar ring galaxies IV.}

 \maketitle

\section{Introduction}  % Sect. 1
A polar-ring galaxy (hereafter referred to as PRG) consists of a
flattened galaxy with an outer ring of gas, dust, and stars rotating
in a plane approximately perpendicular to the central disc.
Kinematically confirmed PRGs with a disc-dominated central
galaxy tend to have wide, extended polar rings, while bulge-dominated
objects show only short, narrow rings (Whitmore 1991; 
Reshetnikov \& Sotnikova 1997).
The PRGs probably represent merger products, and their study
may give us valuable clues about the process and frequency of
merging; their visible environment appears to be similar to that of
normal galaxies (Brocca et al. 1997), which may support long formation 
and evolution times of the rings. 
In addition, measurement of rotation in the two nearly
perpendicular planes of the ring and galaxy provide one of the
few available probes of the three-dimensional shape of galactic gravitational
potentials, and hence the shape of dark and luminous matter distributions.

The catalogue of Whitmore et al. (1990, hereafter PRC) provides us
with over a hundred known polar-ring galaxies and PRG candidates, as well
as a list of possibly related systems, divided into the four main categories 
listed below. The updated number of objects per category listed here is different 
from that tabulated originally in the PRC, as since its publication the following 
4 category B and 4 category C objects  have been promoted to category A: 
B-03 (=IC 1689),  B-17 (=UGC 9562),  B-19 (=AM 2020-504),  B-21 (=ESO603-G21),  C-13 (=NGC 660), 
C-24 (=UGC 4261),  C-27 (=UGC 4385) and C-45 (=NGC 5128); see references in Paper III. 
All of these will be considered members of the A category in our studies
-- note that of these objects only B-19 was observed in our present Parkes survey.

\begin{enumerate}
\item A:  Kinematically-determined Polar-Ring Galaxies (17 objects)
\item B:  Good Candidates for Polar-Ring Galaxies (20 objects)
\item C:  Possible Candidates for Polar-Ring Galaxies (69 objects)
\item D:  Systems Possibly Related to Polar-Ring Galaxies (51 objects)
\end{enumerate}

The scientific goals of our 21-cm \HI\ line surveys of PRGs are to:
\begin{enumerate}
\item Establish redshifts for the objects with previously unknown redshift.
\item Measure the amount of neutral hydrogen in these systems and examine
     its correlations with other observational parameters.
\item Identify objects for subsequent synthesis mapping; \HI\ maps together        
     with optical line studies will show which of the        
     new morphological candidates are true polar-ring galaxies, and
     high-resolution maps will allow dynamical modelling.
\end{enumerate}

% --- Fig. 1 ----------------

\begin{figure*}
\centering
\includegraphics{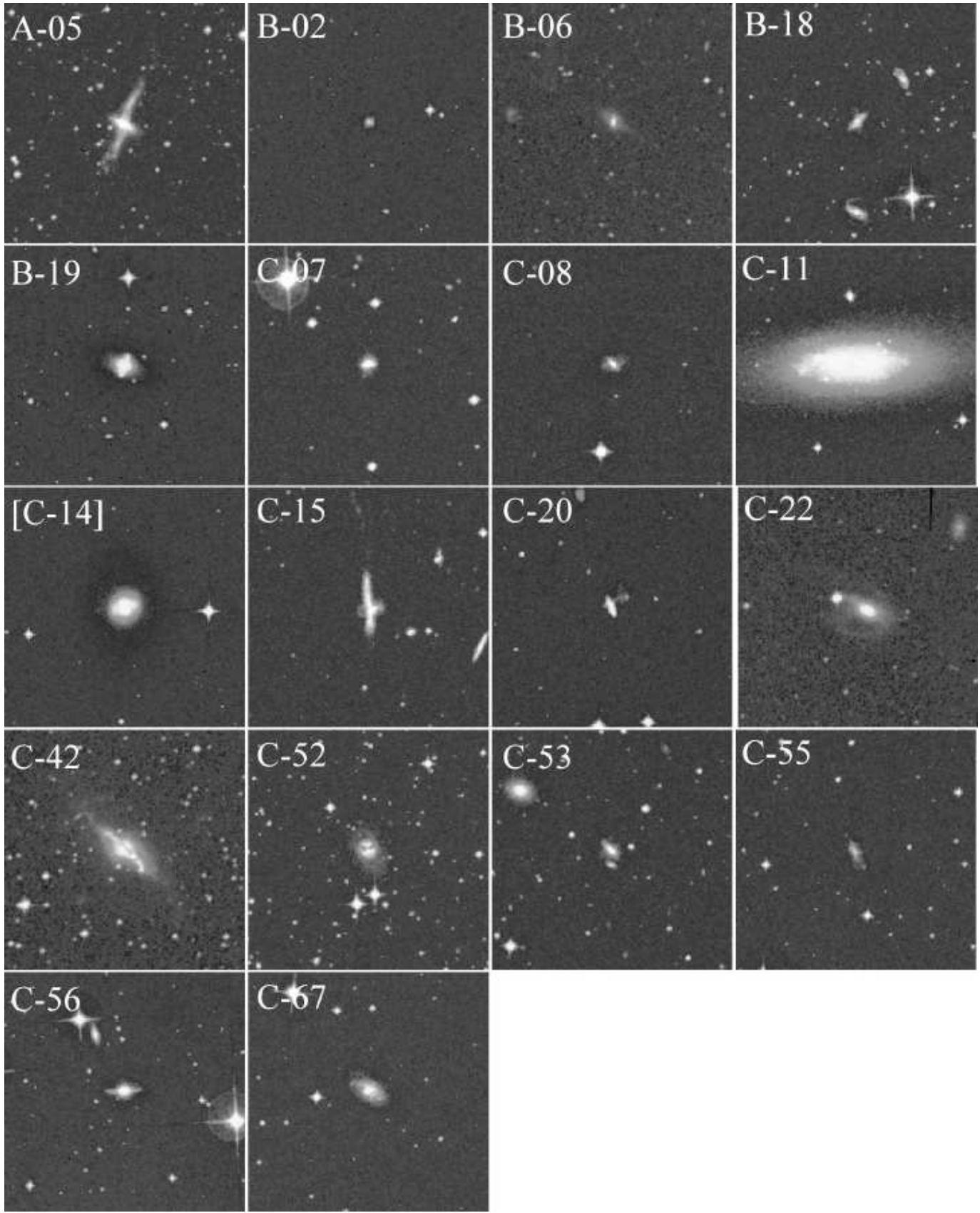}
\caption{Optical Digital Sky Survey images of all 18 PRC objects towards which 
\HI\ line emission was detected at Parkes. Each galaxy is identified by its PRC name. 
Each image has a size of 5$'$$\times$5$'$. For comparison: the telescope's FWHM 
is \am{14}{4}. }
 \end{figure*}

Detailed observations of PRGs in the 21-cm \HI\ line with large radio synthesis
telescopes like the Australia Telescope, VLA or Westerbork, are crucial for an
understanding of the dynamical state of these systems. 
Such mapping measures the distribution and the
velocity field of gas in the rings, both of which are required for
accurate determination of the shape of the dark halo.  Together with
optical absorption-line studies of the central galaxy, rotation in the
ring gas determines whether morphological candidates are true polar
rings. Further, knowledge of the ring mass is necessary to assess
the stability of the rings against differential precession, an important
consideration in estimating the time since its formation.

After having observed a sample of 47 PRC objects in the 21-cm \HI\ line 
with the 140-foot (43-m) Green Bank telescope (Paper I), 44 PRC objects with 
the 100-m Effelsberg telescope (Huchtmeier 1996, Paper II) and 50 PRC objects 
with the 100-m class Nan\c{c}ay telescope (van Driel et al. 2000,
Paper III), we now present observations of 33 PRC objects in the southern
hemisphere with the 64-m Parkes telescope.  
Provisional results of a pilot PRG \HI\ survey made at Parkes in 1992 by a 
member of our team, O.-G. Richter (private communication), 
were refered to in Paper III.
%are summarized in Section 4.1.

In Section 2 of the present paper the sample of PRGs observed in \HI\ at Parkes 
is presented. The observations are described in Section 3, a
a brief discussion of the results is given in Section 4 and the conclusions are
summarized in Section 5..
An analysis of these, and all other available \HI\ data on PRGs and related objects 
will be presented in paper V in these series (van Driel et al.,
in preparation).

\section{The Parkes polar-ring galaxy sample}  % Sect. 2
Listed in Table 1 are basic data for all 33 galaxies observed 
for our PRG \HI\ survey  at Parkes. These data, compiled 
from many sources, by no means form a homogeneous set. 
The sources used, in order of preference, are the online 
Lyon-Meudon Extragalactic Database (LEDA) [http://leda.univ-lyon1.fr], 
the NASA/IPAC Extragalactic Database (NED) [http://nedwww.ipac.caltech.edu] and the PRC.
Generally, the data listed are mean, corrected values from the LEDA database,
while the data in parentheses are uncorrected published values compiled in NED.
Note that all radial velocities listed in this paper are
heliocentric and calculated according to the conventional optical definition 
($V$=c($\lambda$--$\lambda_{0}$)/$\lambda_{0}$).
The total blue magnitudes, $B_T$ are sometimes indicative only, as they 
cannot be assumed to be on the same photometric scale, nor do all represent the 
true total apparent blue magnitude as defined in, e.g., the RC3; sometimes a 
magnitude measured within a single aperture is used instead. 
Neither do the isophotal diameters, $D_{25}$, represent a 
homogeneous set of measurements, as they sometimes refer  to the size of 
the faint polar ring, and sometimes to the size of the brighter equatorial disc.

Of the 33 PRC objects observed at Parkes, 28 were selected using the following criteria: 
(1) South of declination --39$^{\circ}$ (the limit of our \nan\ survey; 
our Green Bank survey reached --45$^{\circ}$), 
(2) from PRC Categories A, B or C , and 
(3) no published \HI\ detection and not detected in the Parkes HIPASS 
Public Data Release spectra [http://www.atnf.csiro.au/research/multibeam/release/]. 
Of these 28, 14 have optical redshifts, 5 of which  (of B-25, B-26, B-27, C-65 and C-72) 
are outside the range of the HIPASS spectra, --1281 to 12,741 \kms.

Besides the 28 objects thus selected, the telescope time allocation also permitted 
us to observe the following 6 objects: three (PRC A-05 = NGC 4650A, C-11 = NGC 625
and C-42 = NGC 4672) that were detected in HIPASS and previously reported 
as detected in the literature, two (C-15 = ESO 199-IG12 and C-55 = ESO 143-G37) that
were detected in HIPAS, and one (C-14 = NGC 979) whose previously reported detection
seems due to radio interference.
For details on the available \HI\ data for all objects we refer to Table 2
and Section 4.1.

\section{Observations}  % 3
The observations with the Parkes 64-m telescope were made in August 2001, 
using about 75 hours of telescope time. The data were obtained with the 13-element
21-cm Multibeam receiver (Staveley-Smith et al. 1996), of which the inner 7
elements only were consecutively pointed towards the target galaxy.
The FWHM of the telescope in the observing mode used at 21-cm wavelength is 
\am{14}{4}.
The autocorrelator was divided into two pairs of cross-polarized receiver banks, 
each with 2048 channels and a 64~MHz bandpass, giving a radial velocity
coverage of about 12,500 \kms\ and a channel separation of 6.6 \kms.
For objects with known redshifts, the centre frequencies of the two banks were generally tuned 
to an \HI\ velocity of 5,500 \kms\ or 15,000 \kms, depending on the redshift. For the 3 objects 
(PRC B-18, C-22 and C-48) with redshifts in the overlapping region, 9,000 to 12,000 \kms, a centre
velocity of 10,000 \kms was used. For objects of unknown redshift, line searches were first made
in the -1000 to 12,000 \kms\ range and, if not detected, in the 8,500-21,500 \kms\ range;
the two rms. noise levels listed in Table 1 for undetected objects refer to these two velocity
search ranges. The observing time was distributed as equally as possible between the 
different line searches, about 1.25 hours per velocity range per object, hence the similarity
in rms noise levels.

\onecolumn

\begin{sidewaystable}[h!]
\centering
\bigskip
{\scriptsize
\begin{tabular}{lllllrrrrrrrrrrr}
\multicolumn{16}{l}{\footnotesize {\bf Table 1.} Basic data for the Parkes polar ring galaxy 
                     \HI\ survey sample.}\\
\smallskip \\
\hline
\vspace{-2 mm} \\
PRC  &  Ident.  & RA & Dec  & $V_{opt}$ & $B_{T}$ & $D_{25}$ & $V_{HI}$ & $W_{50}$ & $W_{20}$ & $I_{HI}$ & rms & 
   D & \LB\ & \MHI\ & \MHILB\ \\
  &  &  &  &  &  &  &  &  &  &  &  &  & [log] & [log] & \\
 &  & \multicolumn{2}{c}{(J2000.0)} & km/s & mag & $'$ & km/s & km/s & km/s & Jy km/s & mJy & 
   Mpc & \LBsun\ & \Msun\ & \MsunLBsun\ \\
\vspace{-2 mm} \\
\hline
\vspace{-2 mm} \\
\multicolumn{16}{l}{Kinematically-determined polar-ring galaxies} \\ 
\vspace{-2 mm} \\
A-05 &  NGC 4650A    & 12 44 48.8 & -40 42 50 &  2808$\pm$97   &  13.91  & 1.5 &  2880$\pm$3   & 221 & 240  & 19.5 & 7.3   &  35.9 &  9.74 &  9.77 & 1.1 \\
B-19 &  AM 2020-504  & 20 23 54.8 & -50 39 05 &  4989$\pm$69   &  15.97  & 1.2 &  5006$\pm$43: & 390 & 650: &  6.1 & 4.4   &  65.5 &  9.44 &  9.44 & 2.2 \\  
\vspace{-2 mm} \\
\multicolumn{16}{l}{Good candidates for polar-ring galaxies} \\ 
\vspace{-2 mm} \\
B-02 &  PGC 419347   & 01 15 28.7 & -54 26 53 &                &  16.83  & 0.4 &  5361$\pm$27 & 98 & 225: &  1.7 & 3.3     &  70.0 &  9.15 & 
 \multicolumn{2}{l}{\it confused?} \\
B-04 &  PGC  145741  & 03 37 57.6 & -48 55 40 &                & (16.61) & 0.6 &              &     &  &  $<$3.6 & 4.5/3.6 &       &       &       & $<$2.4 \\   
B-05 &  PGC  415379  & 03 52 49.2 & -54 49 50 &                &  16.46  & 0.5 &              &     &  &  $<$4.1 & 4.4/4.7 &       &       &       & $<$2.4 \\ 
B-06 &  AM 0442-622  & 04 43 07.6 & -62 19 42 &                &  16.32  & 0.6 &  7165$\pm$27 & 308 & 435 &  2.8 & 3.8     &  92.7 &  9.60 &  9.76 & 1.4 \\ 
B-18 &  AM 1934-563  & 19 38 38.7 & -56 27 30 & 11649$\pm$60   &  15.93  & 0.6 & 11282$\pm$24 & 193 & 308 &  4.1 & 3.3     & 153.8 & 10.19 & 
 \multicolumn{2}{l}{\it confused} \\ 
B-22 &  PGC  13177   & 23 31 54.6 & -40 45 45 &                &  15.11  & 0.8 &              &     &  &  $<$3.9 & 4.5/4.2 &       &       &       & $<$0.65 \\ 
B-25 &  PGC  164989  & 23 51 41.3 & -39 10 31 & (19540$\pm$50) &  16.08  & 0.6 &              &     &  &  $<$6.6 & 4.4     & 260.4 & 10.59 & $<$11.0 & $<$2.7 \\ 
B-26 &  PGC  165022  & 23 53 20.1 & -40 25 53 & (16470)        & (16.74) & 0.7 &              &     &  &  $<$4.6 & 4.4     & 219.3 & 10.18 & $<$10.7 & $<$3.5 \\ 
B-27 &  ESO 293-IG17 & 23 56 25.3 & -39 10 36 & 15300$\pm$60   &  16.17  & 1.0 &              &     &  &  $<$5.4 & 4.5     & 203.8 & 10.34 & $<$10.7 & $<$2.4 \\ 
\vspace{-2 mm} \\
\multicolumn{16}{l}{Possible candidates for polar-ring galaxies} \\
\vspace{-2 mm} \\
C-07 &  ESO 113-IG4  & 01 00 34.2 & -57 44 56 &  3130$\pm$200  &  15.67 & 0.6 &  3587$\pm$29  & 131 & 275: & 2.1 & 4.3     &  46.2 &  9.25 &  9.03 & 0.6  \\  	
C-08 &  ESO 243-IG19 & 01 02 49.2 & -47 11 12 &                &  16.05 & 0.6 &  6438$\pm$40  & 335 & 435 &  3.2 & 5.1     &  84.9 &  9.63 &  9.73 & 1.3   \\   	
C-10 &  ESO 152-IG3  & 01 28 24.8 & -52 38 08 &  3450$\pm$60   &  16.17 & 0.4 &               &     &  &  $<$2.5 & 5.9     &  44.6 &  9.02 & $<$9.1 & $<$1.1 \\
C-11 &  NGC 625      & 01 35 04.2 & -41 26 15 &   391$\pm$44   &  11.61 & 6.3 &   396$\pm$1   &  79 & 112 & 32.1 & 9.9     &   4.6 &  8.87 &  8.20 & 0.2 \\
C-14 &  NGC 979      & 02 31 38.7 & -44 31 28 &                &  14.22 & 1.1 & [5181         & 241 &     &  2.1 & 3.3]    &       &       &       &  $<$0.22   \\ 
C-15 &  ESO 199-IG12 & 03 03 24.2 & -50 29 51 &  7030$\pm$60   &  14.80 & 1.4 &  6873$\pm$18  & 409 & 532 & 10.0 & 5.6     &  89.8 & 10.18 & 10.28 & 1.3 \\       	
C-16 &  AM 0320-495  & 03 21 56.0 & -49 48 03 &                &  16.11 & 0.7 &               &     &  &  $<$4.0 & 3.8/5.1 &       &       &       &  $<$1.7   \\ 
C-20 &  ESO 201-IG26 & 04 15 17.6 & -50 56 43 &  3802$\pm$47   &  15.45 & 0.6 &  3745$\pm$6   & 213 & 225 &  4.1 & 5.2     &  47.7 &  9.37 &  9.35 & 0.95 \\       	
C-21 &  PGC 495219   & 04 16 18.9 & -47 49 14 &                &  16.88 & 0.6 &               &     &  &  $<$3.7 & 4.0/4.2 &       &       &       & $<$3.2 \\
C-22 &  ESO 202-G1   & 04 16 31.1 & -47 50 51 & 10052$\pm$50   &  14.73 & 1.3 &  9780$\pm$30: & 220 & 315: &  1.4 & 3.1    & 128.2 & 10.52 &  
 \multicolumn{2}{l}{\it marginal}  \\ 
C-42 &  NGC 4672     & 12 46 15.4 & -41 42 22 &  3346$\pm$63   &  14.10 & 2.2 &  3202$\pm$6   & 343 & 377 & 12.7 & 4.8     &  46.5 &  9.89 &  9.81 & 0.84 \\
C-48 &  ESO 326-IG6  & 14 11 08.5 & -40 06 22 &  8598$\pm$45   &  15.32 & 0.9 &               &     &  &  $<$5.3 & 5.1     & 111.3 & 10.16 & $<$10.2 & $<$1.1 \\
C-52 &  ESO 232-G4   & 19 22 46.6 & -51 00 08 &  5600:$\pm$200 &  14.81 & 1.3 &  4951$\pm$20  & 320 & 385 &  2.5 & 3.6     &  64.5 &  9.89 & 9.54 & 0.45 \\ 
C-53 &  IC 4982      & 20 20 37.1 & -71 01 45 &                &  15.58 & 0.6 &  6180$\pm$16  & 190 & 235 &  2.9 & 4.6     &  80.0 &  9.77 & 9.63 & 0.74 \\ 
C-55 &  ESO 143-G37  & 20 37 48.0 & -61 44 50 &                &  16.42 & 0.7 &  3265$\pm$9   &  90 & 151 &  4.8 & 5.1     &  42.2 &  8.87 & 9.31 & 2.7  \\ 
C-56 &  PGC  128148  & 20 44 11.6 & -61 59 19 &                &  15.75 & 0.8 &  3335$\pm$24  & 259 & 485 &  5.7 & 4.4     &  42.6 &  9.15 & 9.38 & 1.7 \\ 
C-61 &  PGC  263886  & 21 20 56.6 & -72 20 17 &                &  16.11 & 0.5 &               &     &  &  $<$4.4 & 4.3/5.5 &       &       &      & $<$1.8 \\ 
C-62 &  ESO 236-IG2  & 21 19 46.8 & -52 14 16 &                &  16.35 & 0.5 &               &     &  &  $<$4.1 & 4.2/4.8 &       &       &      & $<$2.1 \\ 
C-65 &  ESO 287-IG50 & 21 41 57.9 & -46 00 38 & 17700$\pm$35   &  15.76 & 0.6 &               &     &  &  $<$7.0 & 4.7     & 235.3 & 10.63 & $<$11.0 & $<$2.1 \\
C-67 &  ESO 75-G55   & 22 06 24.9 & -67 31 03 &                &  14.93 & 1.0 &  3541$\pm$10  & 173 & 193 &  3.3 & 3.0     &  45.2 &  9.53 & 9.20 & 0.47  \\
C-68 &  PGC  127657  & 22 24 14.5 & -66 02 18 &                &  14.73 & 0.9 &               &     &  &  $<$4.6 & 5.0/5.2 &       &       &      & $<$0.54 \\ 
C-72 &  ESO 240-IG16 & 23 44 48.1 & -49 06 41 & 13664$\pm$60   &  15.88 & 0.6 &               &     &  &  $<$4.6 & 4.1     & 179.0 & 10.35 & $<$10.5 & $<$1.6 \\
\vspace{-2 mm} \\
\hline
\vspace{-2 mm} \\
\multicolumn{16}{l}
{Notes: most optical data are mean, corrected values from the LEDA database; data in parentheses 
are literature values, from NED. For the estimated upper limits to} \\
\multicolumn{16}{l}
{the integrated \HI\ line flux, $I_{HI}$, see Sect. 4. A `:' denotes an uncertain value. The line 
detected towards PRC C-14 seems due to nearby ESO 246-G22;} \\ 
\multicolumn{16}{l}
{for details on this object, and on the detections marked as 'confused' or 
`marginal', see Sect. 4.1.} \\
% \hline
\end{tabular}
}
\normalsize
\end{sidewaystable}

\twocolumn

We calibrated, averaged and made a first baseline subtraction in our spectra using the standard 
multibeam receiver spectral line data reduction package available at the Parkes site, GRIDZILLA.
The data were then reformatted, using a routine written by L.S. Staveley-Smith, for further 
reduction with the CLASS spectral line data reduction package.
With CLASS we made a final, third-order polynomial baseline subtraction,  
and determined the global \HI\ line parameters and rms. noise level in the spectra, 
after degrading the velocity resolution to 13.2 \kms\ for most spectra (see Sect. 4).

\section{Results}  % 4
The reduced Parkes 21-cm spectra are shown in Figure 1 for all objects towards which line emission 
was detected, clearly or tentatively. For discussion of possible confusion with other galaxies
in the beam, see Section 4.1 -- we assumed that the emission line detected toward PRC C-14
is due to a nearby galaxy.
Basic optical and \HI\ line parameters of all observed objects are listed in Table 1. For the optical 
data we preferentially used the mean parameters from the LEDA database, or literature values from NED 
if these were not available (for the optical redshift of PRC B-18, see Section 4.1).

Columns in Table 1 are as follows: 
(1) galaxy number in the PRC, 
(2) identifications in other catalogues, 
(3 \& 4) optical centre position, which was used as pointing position for the \HI\ observations, 
(5) optical systemic velocity and its estimated uncertainty, if available, 
(6) the total blue magnitude, uncorrected for Galactic or internal extinction, 
(7) blue major axis diameter at the 25 mag arcsec$^{-2}$ isophotal level, 
(8) centre velocity of the \HI\ profile and its estimated uncertainty (see below), 
(9) width of the \HI\ profile measured at the 50\% level of the peak flux density, 
(10) idem, at the 20\% level, 
(11) integrated \HI\ line flux or an estimated 3$\sigma$ upper limit (see below), 
depending on the blue luminosity of the objects, if known (see text),
(12) rms noise level of the spectrum; if two values are listed for an object, the first refers to 
the $\sim$-1000 to 12,000 \kms\ velocity range and the second to the $\sim$8,500 to 21,500 \kms\ range,
(13) distance, derived preferably using heliocentric systemic \HI\ line velocities, or otherwise 
optical velocities, corrected to the Local Standard of Rest following the prescription of 
Sandage \& Tammann (1981, RSA):
\begin{equation}
V_{LSR} = V_{hel} - 79coslcosb + 296sinlcosb - 36sinb\, [km s^{-1}]
\end{equation}
and assuming a Hubble constant H$_0$=75 \kmsMpc,
(14) blue luminosity,
(15) \HI\ mass, derived straight from the measured integrated \HI\ line 
fluxes without any correction factors, such as for beam filling:
\begin{equation}
M_{HI} = 2.356\, 10^5\, D^2\, I_{HI}\,\, [M_{\odot}]
\end{equation}
(16) ratio of the total \HI\ mass and blue luminosity.

The instrumental velocity resolution of the reduced data is 13.2 \kms, except 
for the narrow profile of C-11 (6.6 \kms) and for the faint detection seen towards C-14 (26.4 \kms).
We estimated the uncertainties, $\sigma_{V_{HI}}$, in the central \HI\ velocities, $V_{HI}$, 
following Fouqu\'e et al. (1990):
\begin{equation}
\sigma_{V_{HI}} = 4 R^{0.5}P_{W}^{0.5}X^{-1}\,\, [km/s]
\end{equation}
where R is the velocity resolution in \kms, $P_{W}$=($W_{20}$--$W_{50}$)/2
in \kms\ and X is the signal-to-noise ratio of a spectrum, which we defined as
the ratio of the peak flux density and the rms noise.
According to Fouqu\'e et al., the uncertainty in the linewidths is 2$\sigma_{V_{HI}}$ for $W_{50}$ 
and 3$\sigma_{V_{HI}}$ for $W_{20}$.

% --- Fig. 2 ----------------

\begin{figure*}
\vspace{-4.5cm}
\centering
\includegraphics{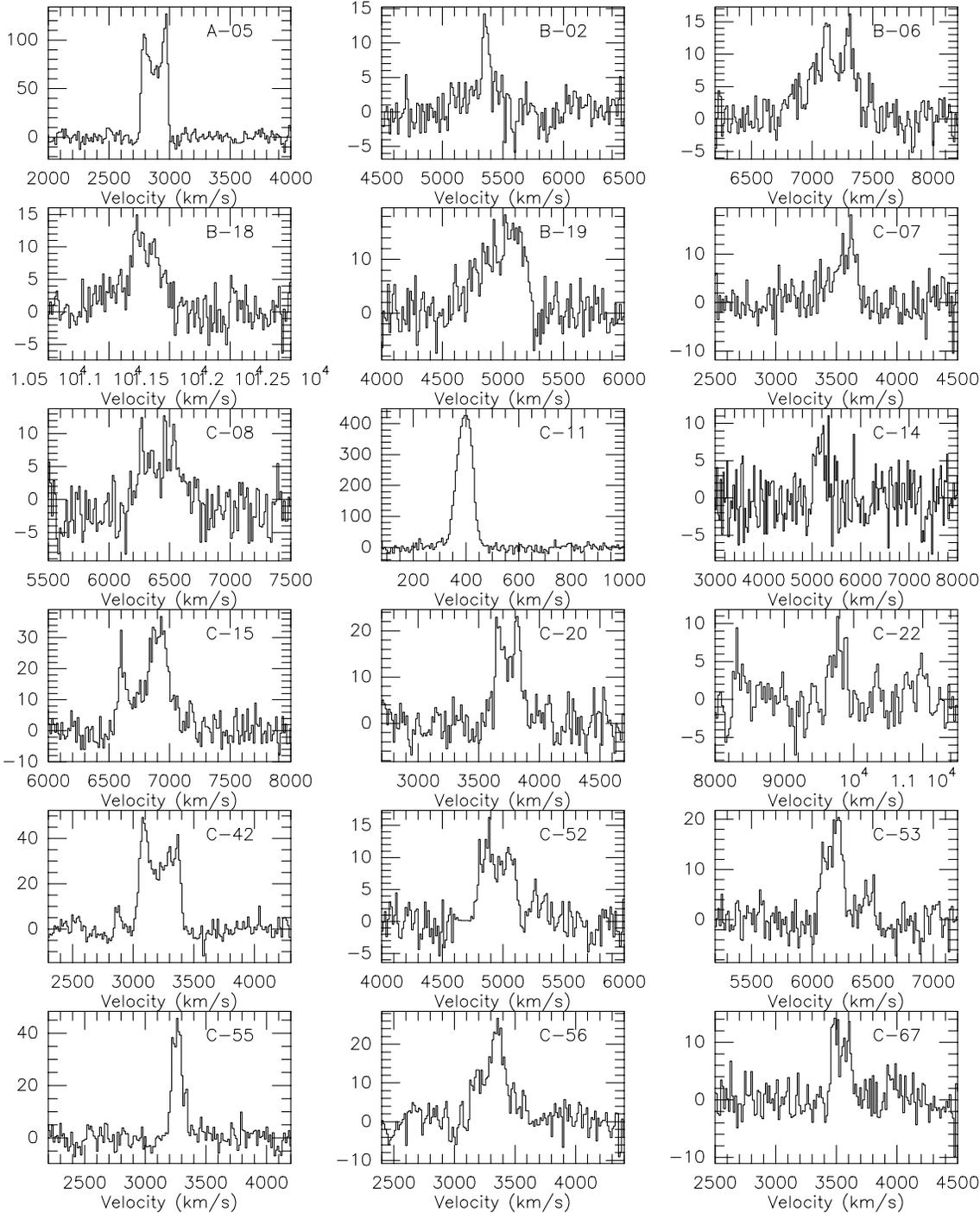}
\vspace{-4.5cm}
\caption{Parkes 21-cm \HI\ line spectra of all clear and marginal detections. 
Each galaxy is identified by its PRC number. The axes are heliocentric 
velocity in \kms, and flux density in mJy. The velocity resolution of the spectra is 
13.2 \kms, except for C-11 (6.6 \kms) and C-14 (26.4 \kms).}
 \end{figure*}

As the \HI\ linewidths of detected PRC objects show an increase with luminosity (van Driel et al., 
in prep.), though with large scatter, the estimated upper limits to the integrated \HI\ line flux, 
$I_{HI}$, are 3 $\sigma$ values for flat-topped profiles with an estimated $W_{20}$ width depending 
on the blue luminosity of the galaxies, consistent with the upper limits listed in Paper III. 
For the 9 objects of unknown redshift, including C-14, a $W_{20}$ linewidth of 300 \kms, 
was assumed, a typical value for the entire sample. For the objects not detected in the 
--1000 to 21,500 \kms\ range, the average rms noise level over this entire range was used to 
estimate the upper limit. The resulting \MHI\ and \MHILB\ upper limits for the objects without
known redshift are obviously only significant if their redshift in fact lies within the velocity 
range searched.

% -------------------------- Table 2 -------------------------------
\begin{table}
\bigskip
{\scriptsize
\begin{tabular}{llrrrrl}
\multicolumn{7}{l}{\footnotesize {\bf Table 2.} Published \HI\ data for the
 Parkes PRG sample.}\\
\smallskip \\
\hline
\vspace{-2 mm} \\     
PRC & Tel. &  $V_{HI}$ & $I_{HI}$ & $W_{50}$ & $W_{20}$ & Ref. \\
    & code &  km/s & Jy km/s & km/s & km/s &  \\
\vspace{-2 mm} \\
\hline 
\vspace{-2 mm} \\
A-05 & P   & 2880 &  19.5 & 221 & 240 & * \\
     & P   & 2910 &       &     & 220 & Ba97 \\
     & G43 & 2909 &  23.2 & 244 &     & Ri94 \\
     & V   & 2910 &  16.0 & 231 & 243 & vG87 \\ 
     & AT  & 2905 &  23.1 & 221 & 240 & Ar97 \\ 
C-11 & P   &  396 &  32.1 &  79 & 112 & * \\
     & P   &  387 &  38.0 &     & 116 & Re82 \\ 
     & P   &  387 &  26.0 &     &     & Bo88 \\  
     & G43 &  404 &  32.4 &     &     & FT81 \\  
     & G43 &  394 &  37.4 & 113 &     & Ri94 \\ 
     & IAR &  370 &  35.1 &  80 & 103 & BM85 \\  
     & V   &  406 &  30.6 &     &     & Co00 \\
C-14 & P   & [5141 &  2.1  & 241  & 241] & * \\
     & P   &     & $<$10.4 &    &     & Ha81 \\   
     & G43 & 4775 &  26.2 & 678: &    & Ri94 \\   
C-42 & P   & 3202 &  12.7 & 343 & 377 & * \\
     & P   & 3289 &  15.6 & 356 & 403 & Aa89 \\ 
     & G43 & 3242 &  7.7  & 401 &     & Ri94 \\  
\vspace{-2 mm} \\
\hline 
\vspace{-2 mm} \\
\multicolumn{7}{l}{Telescope codes:} \\
\vspace{-2 mm} \\
 AT  & \multicolumn{6}{l}{Australia Telescope }  \\ 
 G43 & \multicolumn{6}{l}{Green Bank 43-m } \\ 
 IAR & \multicolumn{6}{l}{I.A.R. 30-m     } \\ 
 P   & \multicolumn{6}{l}{Parkes 64-m     } \\
 V   & \multicolumn{6}{l}{VLA             } \\
\vspace{-2 mm} \\
\multicolumn{7}{l}{References:} \\
\vspace{-2 mm} \\
 Aa89  & \multicolumn{6}{l}{Aaronson et al. (1989)    } \\
 Ar97  & \multicolumn{6}{l}{Arnaboldi et al. (1997)   } \\
 BM85  & \multicolumn{6}{l}{Bajaja \& Martin (1985)   } \\
 Ba97  & \multicolumn{6}{l}{Barnes et al. (1997)      } \\
 Bo88  & \multicolumn{6}{l}{Boiss\'e et al. (1988)    } \\  
 Co00  & \multicolumn{6}{l}{C\^{o}t\'e et al. (2000)  } \\
 FT81  & \multicolumn{6}{l}{Fisher \& Tully (1981)    } \\  
 Ha81  & \multicolumn{6}{l}{Hawarden et al. (1981)    } \\
 Re82  & \multicolumn{6}{l}{Reif et al. (1982)        } \\
 Ri94  & \multicolumn{6}{l}{Richter et al. (1994)     } \\
 vG87  & \multicolumn{6}{l}{van Gorkom et al. (1987)  } \\  
 $\star$ &  \multicolumn{6}{l}{this paper  } \\  
\end{tabular}
}
\end{table}
% -----------------------------------------------------------------

\subsection{Notes on individual objects}  % 4.1
The data on possible companions that may confuse the Parkes spectra
were compiled using the online NED and LEDA databases, searching 
in an area of 22$'$ diameter (i.e., 1.5 times the telescope's FWHM) centered 
on the pointing positions of the telescope (see Table 1). 
The published \HI\ data on our target PRC objects are summarized and compared to our 
present Parkes results in Table 2.
\medskip

{\it Kinematically-determined polar-ring galaxies}
\medskip

{\bf A-05} (= NGC 4650A): This is probably the best-studied of all 
polar ring galaxies, see, e.g., Arnaboldi et al. (1997).
Detected in the present survey and in 4 others,
including two interferometric mappings (see Table 2). The VLA and Australia Telescope 
\HI\ imaging show that our single-dish spectrum is not confused by the two other 
relatively bright, but early-type galaxies in the Parkes beam, 
NGC 4650 ($V_{opt}$=2859 \kms) and NGC 4650B ($V_{opt}$=2515 \kms). 
These early-type galaxies are expected to be gas-poor

{\bf B-19} (= AM 2020-504): This is the best-studied case of a polar ring around an 
elliptical galaxy (Arnaboldi et al. 1993a); the other case in our sample is PRC C-42, 
see below. 
The polar ring shows a gentle warp and is probably stable, giving time to stars to 
form and grow old. 
It has a UV spectrum typical of a starburst galaxy (Arnaboldi et al. 1993b). 
The object was detected in the present survey, at 
$V_{HI}$=5006$\pm$43 \kms. Our profile does not seem to be confused by another 
object in the beam. 
The \HI\ line velocity corresponds well to the mean optical 
value of the PRC object, 4989$\pm$69 \kms\ (LEDA). 

Just outside the telescope's FWHM radius, two galaxies lie at about 8$'$ distance 
from the PRC target: the quite face-on (b/a=0.8) 14.0 mag Sc type spiral ESO 234-G016 
at $V_{opt}$=5215$\pm$82 \kms\ and the 14.6 mag Sbc type spiral ESO 234-G017, of 
unknown redshift. Given the difference in optical velocities (225 \kms) and the range 
of velocities in our \HI\ profile, it seems unlikely that the Parkes profile was 
confused by ESO 234-G016.
\medskip

{\it Good candidates}
\medskip

{\bf B-02} (= PGC 419347): This object does not have a known optical redshift. 
We detected an \HI\ profile towards it, at $V_{HI}$=5361$\pm$27 \kms. 
At \am{7}{3} distance, around the telescope's FWHM radius, 
lies 15.5 mag Sc? type spiral ESO 151-G37, with $V_{opt}$=5452$\pm$60 \kms\ 
(Mathewson \& Ford 1996), 91 \kms\ higher than the \HI\ value. 
If the detection were in fact entirely due to ESO 151-G37, its integrated 
\HI\ line flux would be $\sim$3.4 Jy \kms, at $\sim$50\% beam efficiency, implying an 
\MHILB\ ratio of 0.8 \MsunLBsun. Such a ratio is about 2.5 times the average for
Sc spirals (Roberts \& Haynes 1994) and therefore highly unlikely. 
If, on the other hand, all \HI\ were to be due to B-02 alone, its 
\MHILB\ ratio would be 1.4 \MsunLBsun, a quite common value for an object 
from PRC Category B. A long-slit H$\alpha$ spectrum of ESO 151-G37
(Mathewson \& Ford 1996) shows a maximum rotational velocity, uncorrected for
inclination, of 107 \kms, much larger than half the $W_{50}$ width of our \HI\ spectrum, 
45 \kms. 
In conclusion, it is not very likely that our \HI\ detection may be confused by 
nearby ESO 151-G37.

{\bf B-06} (= AM 0442-622): detected in the present survey, at     
$V_{HI}$=7165 \kms. The only other catalogued galaxy within the telescope's FWHM, 
the low surface brightness object LSBG F118-043, of unknown redshift
and $B_T$ 18.7 mag. Though low surface brightness galaxies can be quite gas-rich, 
it is $\sim$2.4 mag fainter than B-06 in the blue, and hence not a likely candidate 
for confusion with our \HI\ profile. The \MHILB\ ratio of 1.4 \MsunLBsun\ derived
for B-06 assuming that all detected gas resides in it is a quite common value 
among the PRC Category B objects.

{\bf B-18} (= AM 1934-563): detected in the present survey, at 
$V_{HI}$=11282$\pm$24 \kms, with a $W_{50}$ width of 193 \kms. 
Its nuclear H$\alpha$ and $[$N{\sc ii}$]$ emission lines 
(Reshetnikov et al. 2001) indicate fast rotation of the gas in the 
innermost regions, reaching a projected rotational velocity of 
$\sim$230 \kms\ at $\sim$2$''$ from the nucleus, as well as signs
of Sy2 or LINER activity.
Its mean optical systemic velocity listed in LEDA, 11703$\pm$153 \kms, 
did not take into account the latter two of the four following 
published redshifts: 11556$\pm$48 \kms\ (Fisher et al. 1995), 
11842$\pm$120 \kms\ (Allen et al. 1991),
11850$\pm$200 \kms\ (Fairall \& Jones 1991) and
11613$\pm$54 \kms\ (Reshetnikov et al. 2001).
From these three values, we derive a weighted mean value of 
$V_{opt}$=11649$\pm$60 \kms, 367 \kms\ ($\sim$6$\sigma_{V_{opt}}$) 
lower than the \HI\ value, a significantly large difference. 
Three galaxies of similar size and magnitude can be seen
in Figure 1, which appear to form a triplet. 
Besides the PRC object, these are PGC 399718 ($B_T$=15.7 and $D_{25}$=\am{0}{7})
and PGC 400092 ($B_T$=16.2 and $D_{25}$=\am{0}{5}), both of unknown redshift. 
The \MHILB\ ratio of 1.5 \MsunLBsun\ we would derive for B-18 if all
\HI\ were concentrated in it is quite common for a PRC Category B object,
however. 
In conclusion, we cannot exclude that the \HI\ profile may be confused with,
or due to, another member of the triplet.

{\bf B-27} (= ESO 293-IG17): not detected in the present survey.
Our Green Bank \HI\ survey (Paper I) did not cover the optical redshift of the 
object (15,300 \kms), which was published afterwards.
\medskip

{\it Possible PRG candidates}
\medskip

{\bf C-07} (= ESO 113-IG4): an \HI\ line was detected in the present survey, 
at $V_{HI}$=3587$\pm$29 \kms. 
The optical systemic velocity, 3130 \kms\ (from Keel 1985), is 477 \kms\ lower 
than the \HI\ value. Though its uncertainty is listed as 35 \kms\ in LEDA, a value of
200 \kms\ was quoted in Keel's paper - we adopted the latter uncertainty. 
Given the relatively large uncertainties involved, the \HI\ and optical velocities
may be consistent.
No candidates for confusion with our \HI\ spectrum were found within the search area.

{\bf C-08} (= ESO 243-IG19): detected in the present survey, at $V_{HI}$=6438$\pm$40 \kms.
No optical redshift is known of this object. 
No candidates for confusion with our \HI\ spectrum were found within the search area.  

{\bf C-11} (= NGC 625): detected in the present survey and in 6 others (see Table 2),
with good agreement between the global \HI\ line parameters measured at different 
telescopes and in different studies. The galaxy was mapped in \HI\ at the VLA
(C\^{o}t\'e et al. 2000), where a peculiar gas distribution and kinematics was found.
The \HI\ is rotating around the optical major axis, rather than the minor axis, though
its distribution is elongated along the major axis - contrary to the situation in polar rings.
Multiple velocity peaks occur at many places in the VLA data. Consequently no \HI\ rotation
curve could be derived. The ionized gas in the galaxy
also follows complex orbits (Marlowe et al. 1997), which do not match the \HI\ kinematics,
however. Its is a rather red system (B-R) = 0.89), with a smooth morphology reminiscent of 
lenticular galaxies, and strong emission lines, like a blue compact dwarf (Skillman et al. 2002)
According to C\^{o}t\'e et al., the seriously disturbed nature of NGC 625 is due to a
major merger event rather than a case of bad digestion of an accreted \HI\ cloud.

{\bf C-14} (= NGC 979): No optical velocity is known of this object,
and it does not appear to have been detected in \HI. 
A detection with the Green Bank 43-m telescope was reported
in Paper I, with a centre velocity of 4775 \kms, an exceptionally large width
of $W_{20}$=678 \kms\ and a peak flux density of $\sim$30 mJy, while
at Parkes no line signal was detected by Hawarden et al. (1981), with
a published upper limit to the integrated \HI\ line flux of 10.4 \Jykms.
As no trace of the signal seen at Green Bank can be found in our Parkes data, 
which have an rms noise level of 3.3 mJy, it must have been due to interference. 
Our Parkes spectrum shows a marginal \HI\ 
detection at $V_{HI}$=5240 \kms, which is likely to be associated with ESO 246 G-22, 
a 14.4 mag SBc spiral with an optical redshift of 5127$\pm$60 \kms (LEDA), \am{9}{1} 
from the PRC object. We therefore consider PRC C-14 to be undetected in our
survey, and the upper limit to its \MHILB\ ratio of 0.22 \MsunLBsun\ listed
in Table 1 was estimated using a 300 \kms\ linewidth, like for all objects
of unknown redshift.

{\bf C-15} (= ESO199-12): A strong \HI\ line was detected in the present survey, 
with $V_{HI}$=6873$\pm$18 \kms\ and $W_{50}$=409 \kms.
There appears to be a faint detection in the Parkes HIPASS survey data, at $\sim$7000 \kms, 
with a peak flux density of about 30 mJy, like in our spectrum. 
The large uncertainty in the optical velocity of the object, as listed in LEDA,
6727$\pm$358 \kms, is due to one discrepant measurement: 6296$\pm$300 \kms\ (Kirhakos \& Steiner 1990),
compared to the 7033$\pm$35 and 7026$\pm$75: \kms\ from, respectively, Da Costa et al. (1991) and
Chincarini et al. (1984). Therefore, the optical redshift is rather 7030$\pm$60 \kms,
157 \kms\ (i.e., 2.5$\sigma_{V_{opt}}$) higher than the \HI\ value. 
The only other galaxy of significant size ($D_{25}$=\am{1}{2}) within the telescope's FWHM  
is IC 1877, an edge-on Sb:pec spiral of unknown redshift and $B_T$16.2 mag.
It does not seem likely that this 1.4 mag fainter object would cause serious confusion
in the \HI\ spectrum of the target object. The \MHILB\ ratio of 1.25 \MsunLBsun\ we would 
derive if all \HI\ was concentrated in the PRC object is not uncommon for PRC category-C objects.

{\bf C-16} (= AM 0320-495): our Parkes spectrum only shows an off-beam detection of an 
unidentified galaxy, with $V_{HI}$=1023 \kms, $W_{50}$=207 \kms\ and $I_{HI}$=-2.1 \Jykms.
It appears to be a sidelobe-detection of $B_T$ 13.6 mag, Scd spiral IC 1914, at \am{27}{3}
distance from the target object. {\it Observations of IC 1914} at Parkes by 
Longmore et al. (1982) show $V_{HI}$=1037 \kms, $W_{50}$=206 \kms\ and $I_{HI}$=53.0 \Jykms.

{\bf C-20} (= ESO 201-IG26): detected in the present survey, at $V_{HI}$=3745$\pm$6 \kms, 
corresponding well with the optical systemic velocity of  3802$\pm$47 (LEDA). Also in, the 
less sensitive, Parkes HIPASS data a tentative detection is seen at the same velocity.
No candidates for confusion with our \HI\ spectrum were found within the search area:
at \am{2}{4} distance, low surface brightness object LSBG F202-064, 
of unknown redshift and $B_T$ 17.7 mag, does not seem a likely candidate:
though low surface brightness galaxies can be quite gas-rich, 
it is $\sim$2.2 mag fainter than C-20. The \MHILB\ ratio of 0.95 \MsunLBsun\ derived
for C-20 assuming that all detected gas resides in it is a quite common value 
for a PRC Category C objects.

{\bf C-22} (= ESO 202-G1): in our present survey no \HI\ line is seen at the optical velocity, 
10052$\pm$50 \kms\ (LEDA; based on 2 published values), while a tentative line signal is 
seen at $\sim$9780$\pm$30 \kms, with $W_{50}$=220 \kms\ and $I_{HI}$=1.4 \Jykms.
% out: , and a line was reported as detected in our 1992 Parkes pilot survey, 
% out: with $V_{HI}$=9705 \kms, $W_{50}$=180 \kms\ and $I_{HI}$=1.4 \Jykms.
Though no candidates for confusion with our \HI\ spectrum were found within the search area,
we cannot be certain that the object was detected, seen the weakness of the tentative line 
detections and the difference of 272 \kms\ (i.e., 5.5$\sigma_{V_{opt}}$) between the 
well-established optical redshift and our \HI\ value.
We have therefore marked the spectral line as `marginal' in Table 2, and we did not
derive an \HI\ mass from it.

{\bf C-42}  (= NGC 4672): it has an elliptical stellar core rotating perpendicularly to 
the disk and appears to be the end result of the accretion of material in polar orbits
in a disk around a pre-existing oblate spheroid, like in the prototype PRC B-19 
(Sarzi et al. 2000). 
The object was detected in the present survey, at $V_{HI}$=3202$\pm$6 \kms, as well 
as previously at Parkes and at Green Bank (see Table 2) at similar velocities, which are 
all consistent with the mean optical systemic value of 3346$\pm$63 \kms\ (LEDA). 
The integrated line flux  of 7.7 \Jykms\ measured at Green Bank (Paper I) is significantly lower
than the 12.7-15.6 \Jykms\ found at Parkes (Aaronson et al. 1989, and the present paper).
No candidates for confusion with our \HI\ spectrum were found within the search area:
though NGC 4677, a 13.9 mag object at \am{10}{7} distance, has a velocity of 3135$\pm$40 \kms\
(LEDA), similar to that of the PRC object, it is well outside the telescope's FWHM and classified 
as SB0 and therefore expected to be gas-poor. 

{\bf C-48}  (= ESO 326-IG6): not detected in the present survey, nor in our Green Bank 
survey (Paper I), with an almost two times higher rms noise level of 9.5 mJy.

{\bf C-52} (= ESO 232-G4): detected in the present survey, at $V_{HI}$=4951$\pm$20 \kms. 
Its published optical systemic velocities are 5083$\pm$57 \kms\ (Reshetnikov et al. 2001),
based on faint lines, and 5600:$\pm$200 \kms\ (Fairall 1984), a rather uncertain value.
Seen the uncertainties, these values may be consistent with the \HI\ value.
Its nuclear spectrum was tentatively classified as LINER: by Reshetnikov et al. (2001).
No candidates for confusion with our \HI\ spectrum were found within the search area.  

{\bf C-53} (= IC 4982): detected in the present survey, at $V_{HI}$=6180$\pm$16 \kms. 
No object is visible on the DSS at the position
given in the PRC, 20$^h$15$^m$\tis{22}{4}, -71$^{\circ}$11$'$13$''$ (B1950.0); 
the galaxy identified as C-53 on Figure 2 in the PRC is
actually IC 4982, \am{1}{9} NW of the C-53 position. We therefore 
used the optical position of IC 4982 as our pointing centre at Parkes and
listed its optical characteristics in Table 1. 
No candidates for confusion with our \HI\ spectrum were found within the search area:
IC 4985, \am{2}{4} NE of the target object is a 14.8 mag S0 galaxy at a quite
different redshift of $V_{opt}$=4442 \kms (LEDA), and expected to be gas-poor.

{\bf C-55} (= ESO 143-G37): detected in the present survey, at $V_{HI}$=3265$\pm$9 \kms. 
The HIPASS data show an \HI\ detection similar to ours.
No candidates for confusion with our \HI\ spectrum were found within the search area:
no data are available for the small galaxy pair AM2033-620 at \am{6}{2} from the 
target object.

{\bf C-56} (= PGC  128148): detected in the present survey, at $V_{HI}$=6873$\pm$18 \kms.
No candidates for confusion with our \HI\ spectrum were found within the search area.

{\bf C-67} (= ESO 75-G55): detected in the present survey, at $V_{HI}$=6873$\pm$18 \kms. 
No candidates for confusion with our \HI\ spectrum were found within the search area.

\section{Conclusions}
Of the 18 PRC objects towards which \HI\ line emission was noted, 
confusion with one or more galaxies in the telescope beam 
is suspected, though with varying degrees of probability, in three cases 
(PRC B-02, B-18 and C-14), while one weak \HI\ line (of PRC C-22) is 
regarded as a tentative detection only, seen the significant difference between 
the \HI\ and optical velocities. A previously reported detection of
PRC C-14, apparently caused by radio interference, was not reconfirmed. 
Comparing with literature values as well as HIPASS data, the following 
eight objects were detected for the first time in \HI:
PRC B-06, B-19, C-07, C-08, C-53, C-55, C-56 and C-67.
Of the 33 observed galaxies 22 now have known radial velocities, five of which
were determined here for the following objects with previously
unknown redshifts: PRC B-06, C-08, C-53, C-56 and C-67.

\acknowledgements{ 
We are grateful to the ATNF staff at the Parkes observatory and in Epping for 
their assistance with the observations and data reduction. 
The Parkes Telescope is part of the Australia Telescope, which is funded by the 
Commonwealth of Australia for operation as a national facility managed by 
CSIRO. 
This research has made use of the NASA/IPAC Extragalactic Database (NED),   
which is operated by the Jet Propulsion Laboratory, California Institute   
of Technology, under contract with the National Aeronautics and Space      
Administration and the Lyon-Meudon Extragalactic Database (LEDA).
WvD acknowledges the financial support of the ASTE of INSU
for the observations at Parkes.}


\begin{thebibliography}{}
\bibitem[]{}
Aaronson, M., Bothun, G. D., Cornell, M. E., et al. 1989, ApJ, 338, 654 
\bibitem[]{}
Allen, D. A., Norris, R. P., Meadows, V. S., \& Roche, P. F.  1991, MNRAS, 248, 528
\bibitem[]{}
Arnaboldi, M., Capaccioli, M., Cappellaro, E., et al. 1993a, A\&A, 267 21 
\bibitem[]{}
Arnaboldi, M., Barbaro, G. , Buson, L., et al. 1993b, A\&A, 268, 103 
\bibitem[]{}
Arnaboldi, M., Oosterloo, T. A., Combes, F., Freeman, K. C., \& Koribalski, B.,
 1997, AJ, 113, 585
\bibitem[]{} % 
Bajaja, E., \& Martin, M. C. 1985, AJ, 90, 1783
\bibitem[]{} % 
Barnes, L. Staveley-Smith, L., Webster, R. L., \& Walsh W. 1997, MNRAS, 288, 307
\bibitem[]{}
Boiss\'e, P., Dickey, J. M., Kaz\`es, I., \& Bergeron, J. 1988, A\&A 191, 193
% \bibitem[]{}
% Buta, R. 1995, ApJS, 96, 39
\bibitem[]{}
Brocca,  C., Bettoni, D., \& Galletta, G. 1997, A\&A, 326, 907
\bibitem[]{}
Chincarini, G.,Tarenghi, M., Sol, H., et al. 1984, A\&AS, 57, 1
\bibitem[]{}
C\^{o}t\'e, S., Carignan, C., \& Freeman, K. C. 2000, AJ, 120, 3027
\bibitem[]{}
Da Costa, L. N., Pellegrini, P. S., Davis, M., et al. 1991, ApJS, 75, 935
\bibitem[]{}
Fairall, A.P. 1984, MNRAS, 210, 69
\bibitem[]{}
Fairall, A., \& Jones, A. 1991, Southern Redshifts Catalogue,
 Pub. of the Dept. of Astronomy, Univ. of Cape Town, num. 11  
% \bibitem[]{}
% Faundez-Abans, M., \& de Oliveira-Abans, M. 1997, A\&AS, 129, 357
\bibitem[]{} % 
Fisher, J. R., \& Tully, R. B. 1981, ApJS, 47, 139
\bibitem[]{}
Fisher, J. R., Huchra, J. P., Strauss, M. A., et al. 1995, ApJS, 100, 69
\bibitem[]{}
Fouqu\'e, P., Bottinelli, L., Durand, N., Gouguenheim L., \& Paturel. G., 1990, 
 A\&AS, 86, 473
\bibitem[]{}
Hawarden, T. G., Longmore, A. J., Goss, W. M., Mebold, U., \& Tritton, S. B. 1981, 
 MNRAS, 196, 175
\bibitem[]{}
Huchtmeier, W. 1997, A\&A, 319, 401 (Paper II)
\bibitem[]{}
Keel W. C. 1985, AJ, 90, 2207
\bibitem[]{}
Kirhakos, S. D., \& Steiner, J. E. 1990, AJ, 99, 1722
\bibitem[]{}
Krumm, N., \& Salpeter, E. E. 1976, ApJ, 208, L7 
\bibitem[]{}
Longmore, A. J., Hawarden, T. G., Goss, W. M., Mebold, U., Tritton, S. B. 1982, 
 MNRAS 200, 325
\bibitem[]{}
Marlowe, A., Meurer, G., Heckman, T., \& Schommer, R. 1997, ApJS, 112, 285
\bibitem[]{}
Mathewson, D. S., \& Ford, V. L. 1996, ApJS, 107, 97
\bibitem[]{}
Reif, K., Mebold, U., Goss, W. M., van Woerden, H., \& Siegman, B. 1982, A\&AS, 50, 451
\bibitem[]{}
Reshetnikov, V. P., \& Sotnikova, N. 1997, A\&A, 325, 933
\bibitem[]{}
Reshetnikov, V. P., Fa\'undez-Abans, M., \& de Oliveira-Abans, M. 2001, MNRAS, 322, 689
\bibitem[]{}
Richter, O.-G., Sackett, P. D., \& Sparke, L. S. 1994, AJ, 107, 99 (Paper I)
\bibitem[]{}
Roberts, M. S., \& Haynes, M. P. 1994, ARA\&A 32, 115
\bibitem[]{}
Sandage, A., \& Tammann, R. A. 1981, A Revised Shapley Ames-Catalog of Bright
 Galaxies, Carnegie Inst. of Washington (RSA)
\bibitem[]{}
Sarzi, M., Corsini, E. M., Pizzella, A., et al. 2000, A\&A, 360, 439
\bibitem[]{}
Skillman, E., C\^{o}t\'e, S., \& Miller, B. 2002, AJ, in press
\bibitem[]{}
Staveley-Smith, L., Wilson, W. E., Bird, T. S., et al. 1996, PASA, 13, 243
\bibitem[]{}
van Driel, W., Arnaboldi, M., Combes, F., \& Sparke, L. S., 2000, 
 A\&A,  141, 385 (Paper III)
\bibitem[]{}
van Gorkom, J. H., Schechter, P., \& Kristian, J. 1987, ApJ, 314, 457
\bibitem[]{}
Whitmore, B. C., Lucas, R. A., McElroy, D. B., et al. 1990, AJ, 100, 1489 (PRC)
\bibitem[]{}
Whitmore, B. C. 1991, in: Warped Disks and Inclined Rings around Galaxies, eds. 
 S. Casertano, P. Sackett and F. Briggs (Cambridge University press: Cambridge), 
 p. 60
\end{thebibliography}
\end{document}